\documentclass[aps,prb,reprint,superscriptaddress,showpacs,showkeys,preprintnumbers,%reprint,
nofootinbib,
amsmath,amssymb,
aps,
prb,
]{revtex4-2}

\usepackage{graphicx}% Include figure files
\usepackage{dcolumn}% Align table columns on decimal point
\usepackage{bm}% bold math
\usepackage{mathrsfs}
\usepackage[T1]{fontenc}
\usepackage{scalerel}
\usepackage{siunitx}
\usepackage{physics}
\AtBeginDocument{\RenewCommandCopy\qty\SI}
\DeclareSIUnit\rydberg{\text{Ry}}
\DeclareSIUnit{\au}{a.u.}
\usepackage[normalem]{ulem} 
\usepackage[dvipsnames]{xcolor}
\usepackage{hyperref}
\hypersetup{colorlinks=true, linkcolor=blue, citecolor=blue}
\begin{document}

\title{Granular Superconductivity in La$_{2}$PrNi$_{2}$O$_{7-\delta}$ Thin Films}

\author{Ziao Han}
\affiliation{Beijing National Laboratory for Condensed Matter Physics, Institute of Physics, Chinese Academy of Sciences, Beijing 100190, China}
\affiliation{University of Chinese Academy of Sciences, Beijing 100049, China}

\author{Lifen~Xiang }
%\thanks{These authors contributed equally to this work.}
\affiliation{Beijing National Laboratory for Condensed Matter Physics, Institute of Physics, Chinese Academy of Sciences, Beijing 100190, China}

\author{X.J.~Zhou}
\email{XJZhou@iphy.ac.cn}
\affiliation{Beijing National Laboratory for Condensed Matter Physics, Institute of Physics, Chinese Academy of Sciences, Beijing 100190, China}
\affiliation{University of Chinese Academy of Sciences, Beijing 100049, China}
\affiliation{Songshan Lake Materials Laboratory, Dongguan, China}

\author{Zhihai~Zhu}
\email{zzh@iphy.ac.cn}
\affiliation{Beijing National Laboratory for Condensed Matter Physics, Institute of Physics, Chinese Academy of Sciences, Beijing 100190, China}
\affiliation{Songshan Lake Materials Laboratory, Dongguan, China}

\date{\today}

\begin{abstract}

Superconductivity realized in bilayer nickelate thin films enables direct spectroscopic and transport studies at ambient pressure. However, a persistent two-step resistive transition remains a major barrier to achieving optimal superconducting properties. Here, we show that the two-step transition in La$_2$PrNi$_2$O$_{7-\delta}$ thin films originates from the granular nature of superconductivity, specifically, the coexistence of two distinct superconducting grain phases coupled by a Josephson junction network. A secondary, lower-temperature transition appears in the $R(T)$ curve, even when residual resistance becomes vanishingly small near 30 K. This two-step behavior significantly lowers the zero-resistance transition temperature, $T_{c, zero}$$\approx$ 10 K, and limits advanced spectroscopic studies. Our findings reveal the microscopic mechanism underlying the two-step transition in thin films and underscore the need for improved oxygen homogeneity to achieve bulk superconductivity in this system.

\end{abstract}

\maketitle
\section{Introduction}
The discovery of superconductivity in bulk La$_{3}$Ni$_{2}$O$_{7}$ under 14 GPa pressure, with an onset transition temperature $T_{c, onset}$ of approximately 80 K, has attracted considerable attention to the broad family Ruddlesden–Popper nickelates with varying number of NiO$_2$ layers\cite{sun_signatures_2023,houhou_emergence_2023,zhang_hightemperature_2024,wang_pressureinduced_2024,li_bulk_2026,wang_bulk_2024,wangwang_normal_2024,zhu_superconductivity_2024,shi_pressure_2025,zhang_bulk_2025,gao_preparation_2021,li_superconducting_2020a,liSuperconductivityInfinitelayerNickelate2019b,oppliger_discovery_2025}. More recently, superconductivity has also been induced in La$_{3}$Ni$_{2}$O$_{7}$ thin films via epitaxial growth on SrLaAlO$_{4}$ (SLAO) substrates, where compressive strain leads to a $T_{c, onset}$ exceeding 40 K and a zero-resistance $T_{c, zero}$ in the range of 2.8 to 5.5 K\cite{ko_signatures_2025a,xiang_stabilizing_2026}. Furthermore, partial substitution of La with other rare-earth elements, such as Pr or Sm, or both, can raise $T_{c, onset}$ to 50–60 K, with $T_{c, zero}$ reaching up to 37 K\cite{liu_superconductivity_2025,zhou_ambientpressure_2025a,zhou_superconductivity_2026,hao_superconductivity_2025,shi_critical_2026}. The achievement of high $T_c$ ambient-pressure superconductivity in thin films presents an exciting opportunity to investigate the mechanisms of high-temperature superconductivity in bilayer nickelates\cite{li_angleresolved_2025,luo_bilayer_2023,wang_electronic_2025,sun_observation_2025,nie_superconductivity_2026,shen_nodeless_2025,ren_resolving_2025a,bhatt_structural_2026,zhao_pressureenhanced_2026}.

In thin films of bilayer nickelates, the frequently observed broad, two-step superconducting transitions point to phase inhomogeneity, which remains a major barrier to achieving higher zero-resistance temperatures $T_{c, zero}$. To address this, several strategies have been employed to produce high-quality superconducting films, including isovalent doping with Pr and Sm to suppress competing Ruddlesden-Popper phases, optimizing growth conditions for better crystallinity, and precise ozone annealing to control the oxygen content\cite{liu_superconductivity_2025,hao_superconductivity_2025,lyu_preparation_2025}. Further studies have revealed that lower-temperature transition exhibits Berezinskii-Kosterlitz-Thouless (BKT) behavior, indicative of two-dimensional superconductivity\cite{zhou_ambientpressure_2025a,saito_highly_2016,reyren_superconducting_2007,eley_approaching_2012}. Additionally, phenomena such as hysteretic magnetoresistance and slow resistance relaxation during this transition have been reported, suggesting the emergence of a possible spin-glass phase\cite{ji_timereversal_2026}. This implies a coexistence of superconductivity and spin-glass order, which differs from cuprates but is reminiscent of effects seen in infinite-layer nickelates\cite{saykin_spinglass_2025}. Nevertheless, the microscopic origin of the two-step transition—whether stemming from intrinsic fluctuations, oxygen inhomogeneity, or local phase separation—remains unresolved.

In this Letter, we investigate typical bilayer nickelate films exhibiting pronounced two-step superconducting transitions, systematically exploring their behavior under various ozone annealing conditions and applied magnetic fields. We observe significant hysteresis in the magnetoresistance, suggesting that superconductivity in these La$_2$PrNi$_2$O$_{7-\delta}$ thin films is likely granular in nature. Importantly, our results do not support the broken of time-reversal symmetry, in contrast to recent reports\cite{ji_timereversal_2026}. This discrepancy indicates that spin-glass behavior may depend on specific sample conditions rather than being a universal feature. Our findings point to oxygen inhomogeneity as the primary cause of the two-step transition and underscore the urgent need for methods to eliminate such inhomogeneity to advance the understanding of superconductivity in these systems.

\section{Experiment}
Thin films of La$_{2}$PrNi$_{2}$O$_{7}$ were grown on SLAO(001) substrates (5×5 mm, PrMat Corporation) using pulsed laser deposition (PLD) with a 248-nm KrF excimer laser (COMPex 201, Coherent). During growth, the substrate temperature was maintained at 680 $^\circ$C under an oxygen partial pressure of 150 mTorr. The laser beam size was about 7 mm$^{2}$, achieved with an aperture. The pulse energy of the laser was set to 700 mJ/cm$^{2}$ for the growth of La$_{2}$PrNi$_{2}$O$_{7}$. The laser frequency was set to 4 Hz. After deposition, the films were cooled to room temperature at 5 $^\circ$C/min under the same oxygen partial pressure. The as-grown films of La$_{2}$PrNi$_{2}$O$_{7}$ were then annealed in ozone using a procedure similar to that described in \cite{xiang_stabilizing_2026}.

The superconducting transition temperature was measured using electrical transport on a Quantum Design Physical Property Measurement System (PPMS) with a standard four-probe setup. Cross-sectional specimens for scanning transmission electron microscopy (STEM) were prepared using focused ion beam (FIB) techniques (Helios 600i). High-angle annular dark-field (HAADF) imaging was performed on an ARM-200F microscope (JEOL, Japan) operated at 200 kV, equipped with a CEOS Cs corrector (CEOS GmbH, Germany).

\section{Results and discussion}

\begin{figure*}[htb] \centering
\includegraphics[width=0.7\textwidth]{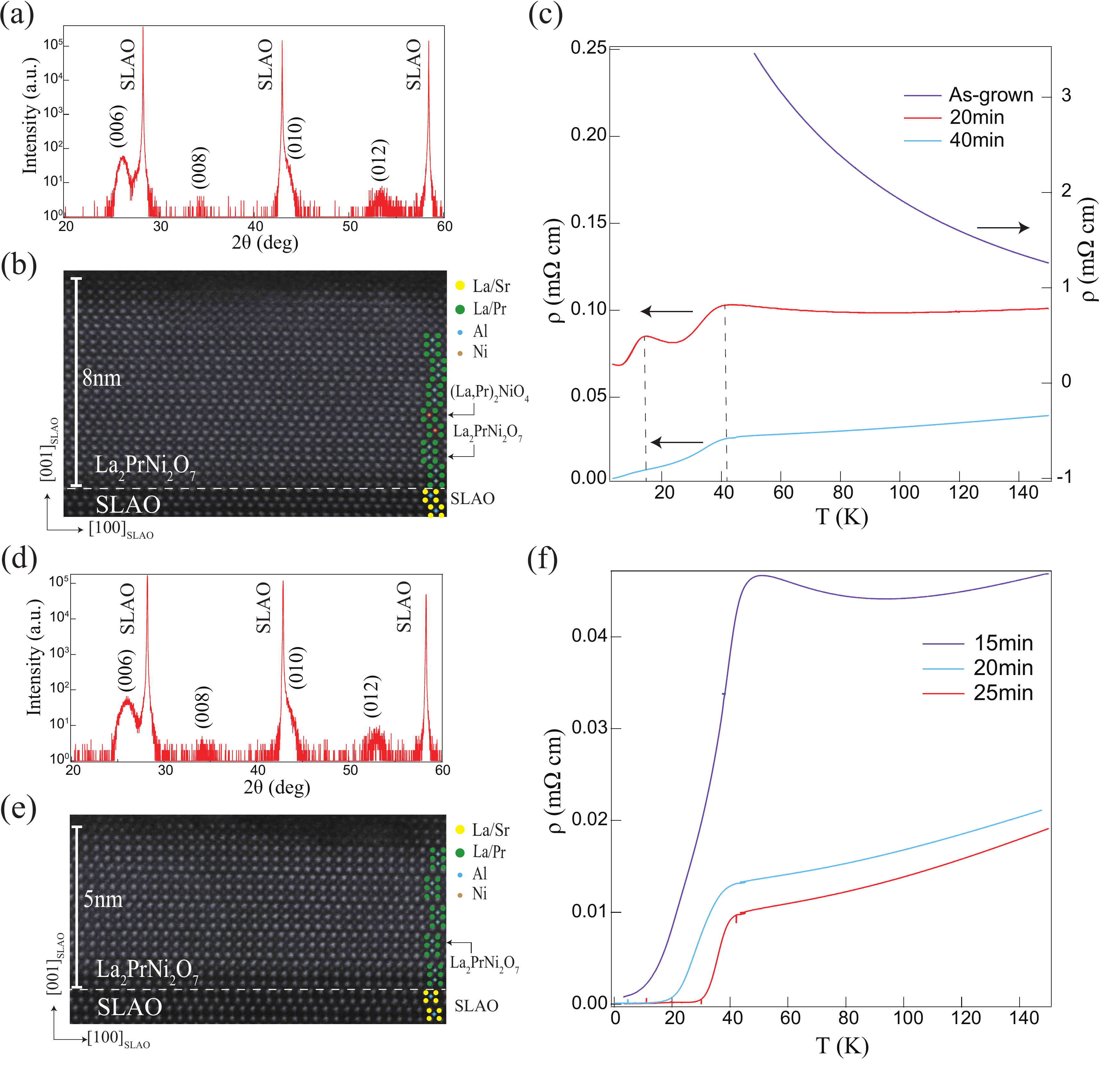}
\caption{{\textbf{Structural characterization and ozone annealing optimization}.} (a) XRD $\theta$–$2\theta$ scans of a typical La$_{2}$PrNiO$_{7}$ film ( Film A, thickness: 8nm). (b) Annular dark-field–STEM image of Film A on SLAO(001) along the [100] zone axis. Across most of the field of view, part of the La$_{2}$PrNi$_{2}$O$_{7}$ structure is replaced by the (La, Pr)$_{2}$NiO$_{4}$ phase. Atomic models are overlaid on the right sides of the image. Blue and red square octahedra represent the bilayer and monolayer structures, respectively. Atom species are color-coded as indicated in the legend. (c) $\rho(T)$ curves of Film A under different ozone annealing times. Annealing conditions are color-coded as shown in the legend. (d) XRD $\theta$–2$\theta$ scans of a typical La$_{2}$PrNi$_{2}$O$_{7}$ film (Film B, thickness: 5 nm). (e) Annular dark-field STEM image of La$_{2}$PrNi$_{2}$O$_{7}$ Film B on SLAO(001) along the [100] zone axis, showing the crystal structure is entirely the bilayer structure of La$_{2}$PrNi$_{2}$O$_{7}$. (f) $\rho(T)$ curves of Film B under different ozone annealing time durations.}
\label{figure1}
\end{figure*}
In Fig.\ref{figure1}, we present the structural characterization and optimization of ozone annealing for two typical La$_{2}$PrNiO$_{7}$ thin films. Fig.\ref{figure1}a shows the out-of-plane XRD pattern from Film A (8-nm-thick), which has a weak (0 0 8) diffraction peak, indicating relatively poor crystallinity and possible structural disorder. 

The annular dark-field STEM image shown in Fig.\ref{figure1}b reveals the local structural variations of Film A: across most of the field of view, parts of the target phase of bilayer La$_{2}$PrNi$_{2}$O$_{7}$ are replaced by (La, Pr)$_{2}$NiO$_{4}$ monolayer intergrowths. The $\rho(T)$ curve of this sample (Fig.\ref{figure1}c) displays two notable features: first, an evident two-step transition with a primary transition at higher temperature and a secondary transition at lower temperature, and second, a clear resistance upturn before both transitions. With increasing ozone annealing time, the upturn in resistance gradually disappears, and the two-step transition is partially suppressed. Conversely, Film B, possessing a thickness of 5 nm, exhibits a more pronounced (0 0 8) diffraction peak in its out-of-plane XRD pattern (Fig.\ref{figure1}d), signifying enhanced crystallinity. The STEM image (Fig.\ref{figure1}e) confirms a relatively uniform crystal structure comprised entirely of the intact La$_2$PrNi$_2$O$_7$ bilayer phase, with no evidence of monolayer intergrowths. The resistivity-temperature curve (Fig.\ref{figure1}f) displays the superconducting transitions around 40 K. However, the two-step transition is much less noticeable than that of Film A (see also Fig.\ref{figure3} below). These results show that structural integrity greatly affects the superconducting behavior. Film A, with its structural disorder and (La, Pr)$_{2}$NiO$_{4}$ monolayer intergrowths, exhibits more prominent two-step transitions. Additionally, the increase in resistance observed before both transitions in Film A provides important experimental evidence for the possible coexistence of two distinct superconducting phases within the films.

\begin{figure*}[htb] \centering
	\includegraphics[width=0.95\textwidth]{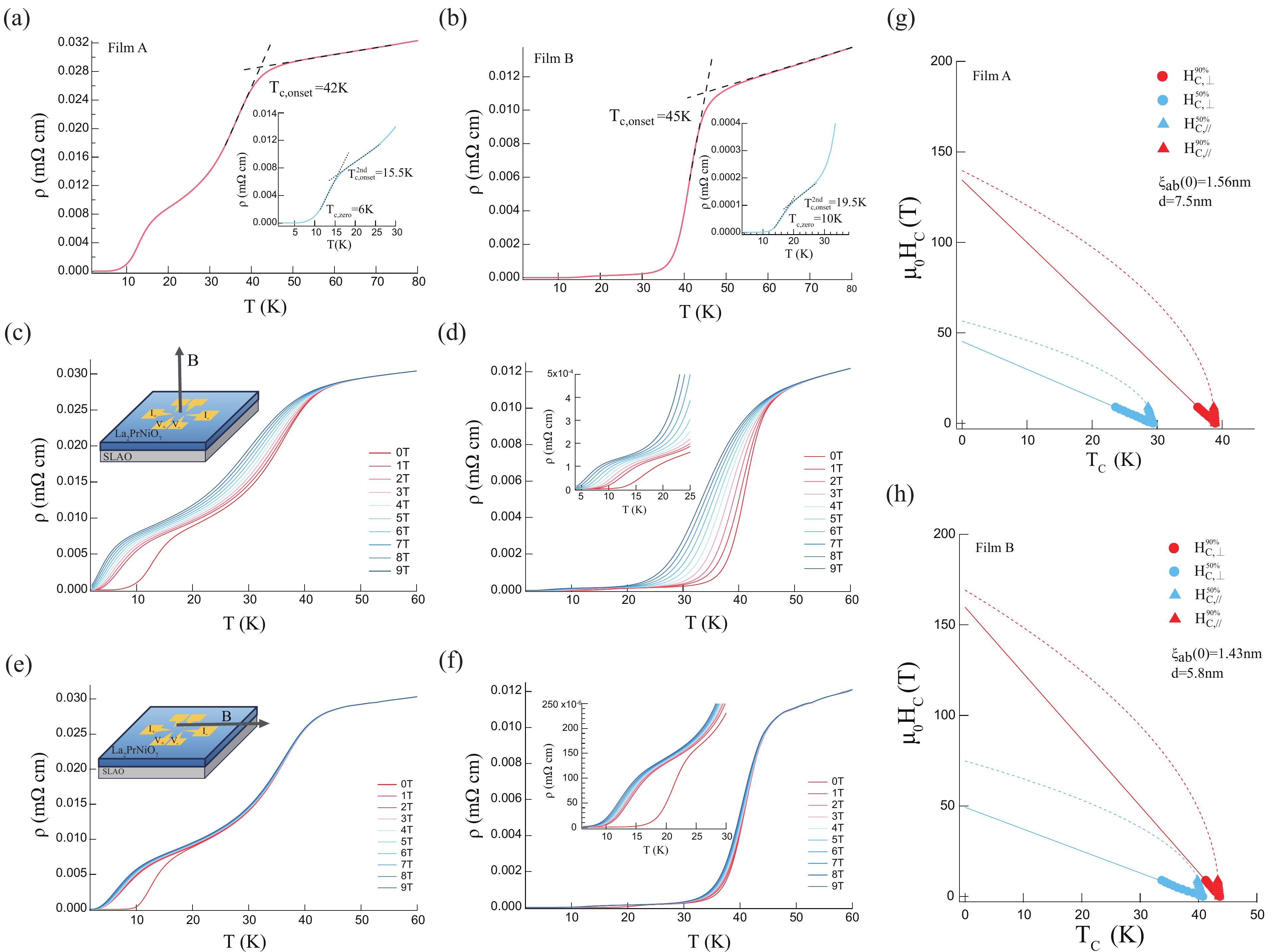}
	\caption{ \textbf{Superconductivity in bilayer nickelate La$_{2}$PrNi$_{2}$O$_{7}$ thin films.} (a, b) Temperature-dependent sheet resistivity $\rho(T)$ curves for Film A and Film B. Insets show the temperatures of the second transition and the onset of zero resistance. (c, d) $\rho(T)$ under various magnetic fields applied perpendicular to the thin film, measured on Film A (c) and Film B (d). Insets illustrate the direction of the applied magnetic field. (e, f) $\rho(T)$ under various magnetic fields applied parallel to the thin film, measured on Film A (e) and Film B (f). (g, h) Upper critical fields extracted using $T_{c,90\%}$ and $T_{c,50\%}$, represented by solid circles and triangles, respectively. Solid lines are Ginzburg–Landau fits.} 
	\label{figure2}
\end{figure*}

We further show in Fig.\ref{figure2} the electrical transport properties of the superconducting thin films A and B, along with their responses to magnetic fields. Using the standard four-point probe method within a Quantum Design Physical Property Measurement System, we systematically studied the transport behavior under magnetic fields applied both parallel and perpendicular to the film $ab$-plane, which is parallel to the substrate. As shown in the figure, Film A has a superconducting onset temperature $T_{c,\text{onset}}$  of $\sim$ 42 K, a secondary transition onset $ T_{c,\text{onset}}^{\text{2nd}} $  of 15.5 K, and a zero-resistance temperature $T_{c,\text{zero}}$  of $\sim$ 6 K (Fig.\ref{figure2}a). For film B, the corresponding characteristic temperatures are $T_{c,\text{onset}} =45 K$, $ T_{c,\text{onset}}^{\text{2nd}}  =19.5 K$, and $T_{c,\text{zero}} =10 K$ (Fig.2b). Under applied magnetic fields, both samples show similar behavior. The resistivity is significantly suppressed with increasing field strength (Fig.2c-f). Notably, the secondary transition is more responsive to relatively weak magnetic fields, a key characteristic of granular superconductors that will be discussed in detail later. The upper critical fields for perpendicular ($H_{c,\perp}$) and parallel ($H_{c,\parallel}$) directions were determined from the temperatures at which the resistivity drops to $90\%$ and $50\%$ of the normal state value (Fig.\ref{figure2}g-h). These data were fitted using the Ginzburg-Landau formulas:
\begin{equation}
	H_{c,\perp}(T) = \frac{\phi_0}{2\pi\xi^2_{ab}}(1-T/T_c)
\end{equation}
\begin{equation}
	H_{c,\parallel}(T) = \frac{\sqrt{12} \phi_0}{2\pi \xi_{ab}(0)d} \left(1 - \frac{T}{T_c}\right)^{\frac{1}{2}}
\end{equation}
where $\phi_0$ is the magnetic flux quantum, $\xi_{ab}(0)$ is the zero-temperature in-plane coherence length, and $d$ represents the thickness of the superconducting layer. For Film A, $H_{c,\perp}^{90\%} \approx 134.4\,\mathrm{T}$, $ H_{c,\perp}^{50\%} \approx 45.3\,\mathrm{T}$, $ H_{c,\parallel}^{90\%} \approx 139.6\,\mathrm{T}$,  and $H_{c,\parallel}^{50\%} \approx 56.5\,\mathrm{T}$. The fitting yields an in-plane Ginzburg-Landau coherence length $\xi_{GL}(0)$ of approximately 1.56 nm and a superconducting layer thickness of $\sim$ 7.5 nm. This thickness matches the total film thickness measured by X-ray reflectivity (8 nm) and TEM (8 nm). For Film B, $H_{c,\perp}^{90\%} \approx 159.8\,\mathrm{T}$, $H_{c,\perp}^{50\%} \approx 49.2\,\mathrm{T}$, $H_{c,\parallel}^{90\%} \approx 169.2\,\mathrm{T}$, and $ H_{c,\parallel}^{50\%} \approx 74.8\,\mathrm{T}$. The fitting yields an in-plane coherence length of $\sim$ 1.43 nm and a superconducting layer thickness of $\sim$ 5.8 nm, consistent with a total thickness of $\sim$ 5 nm.

Fig.\ref{figure3} presents the magnetoresistance response of a typical La$_{2}$PrNi$_{2}$O$_{7}$ film as a function of out-of-plane magnetic field sweeping at various characteristic temperatures. Pronounced hysteresis in the magnetoresistance is observed, along with a subtle fine splitting structure near the resistance minimum (Fig.\ref{figure3}b-g). This closely resembles the typical features reported in the granular superconductor YBa$_{2}$Cu$_{3}$O$_{7-\delta}$ \cite{shifang_behaviour_1988,balaev_correlation_2014}. The arrows indicate the field sweep directions: from 1 T to -1 T (blue lines) and from -1 T to 1 T (red lines). Throughout the hysteresis loop, the resistance on the descending field branch is consistently lower than that on the ascending branch at the same external field value—a characteristic signature of weak-link regions in granular superconductors\cite{balaev_correlation_2014,balaev_magnetoresistance_2007,ji_magneticfielddependent_1993}. 

\begin{figure*}[htb] \centering
 \includegraphics[width=0.9\textwidth]{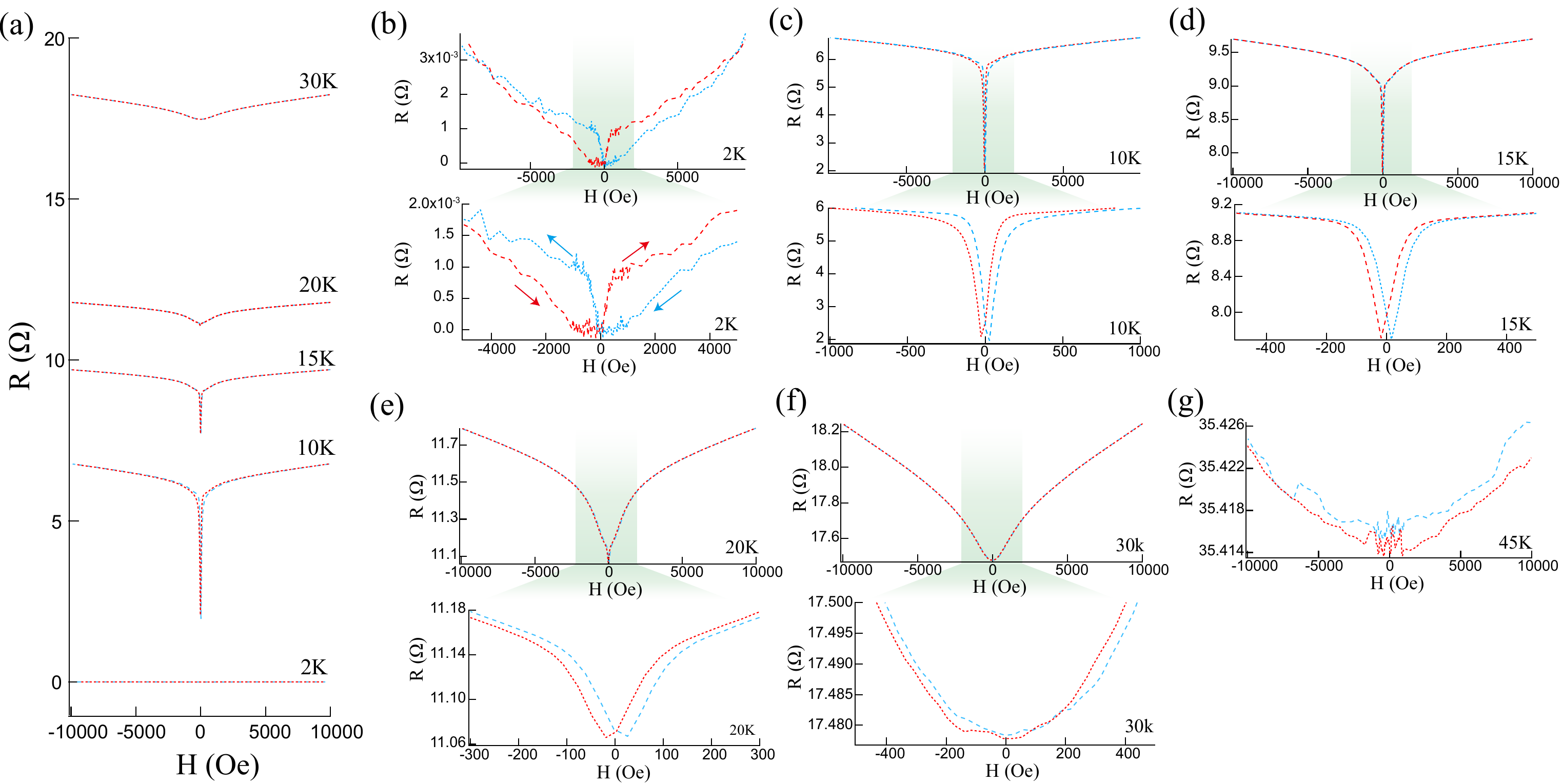}
\caption{\textbf{Hysteretic out-of-plane magnetic field dependence of magnetoresistance $R(H)$ for a typical La$_2$PrNi$_2$O$_7$ thin film sample at several representative temperatures.} (a) $R(H)$ curves measured at temperatures below $T_{c,\text{zero}}$, between $T_{c,\text{zero}}$ and $T_{c,\text{onset}}^{\text{2nd}}$, around $T_{c,\text{onset}}^{\text{2nd}}$, and between $T_{c,\text{onset}}^{\text{2nd}}$ and $T_{c,\text{onset}}$. All $R(H)$ curves are measured with $H_{\perp}$ swept from 1 T to –1 T (blue curves) and from –1 T to 1 T (red curves). (b–g) Enlarged sections of the $R(H)$ curves highlighting finer details. Arrows indicate the direction of the external field $H$.}

\label{figure3}
\end{figure*}
\begin{figure}[htbp] \centering
\includegraphics[width=0.48\textwidth]{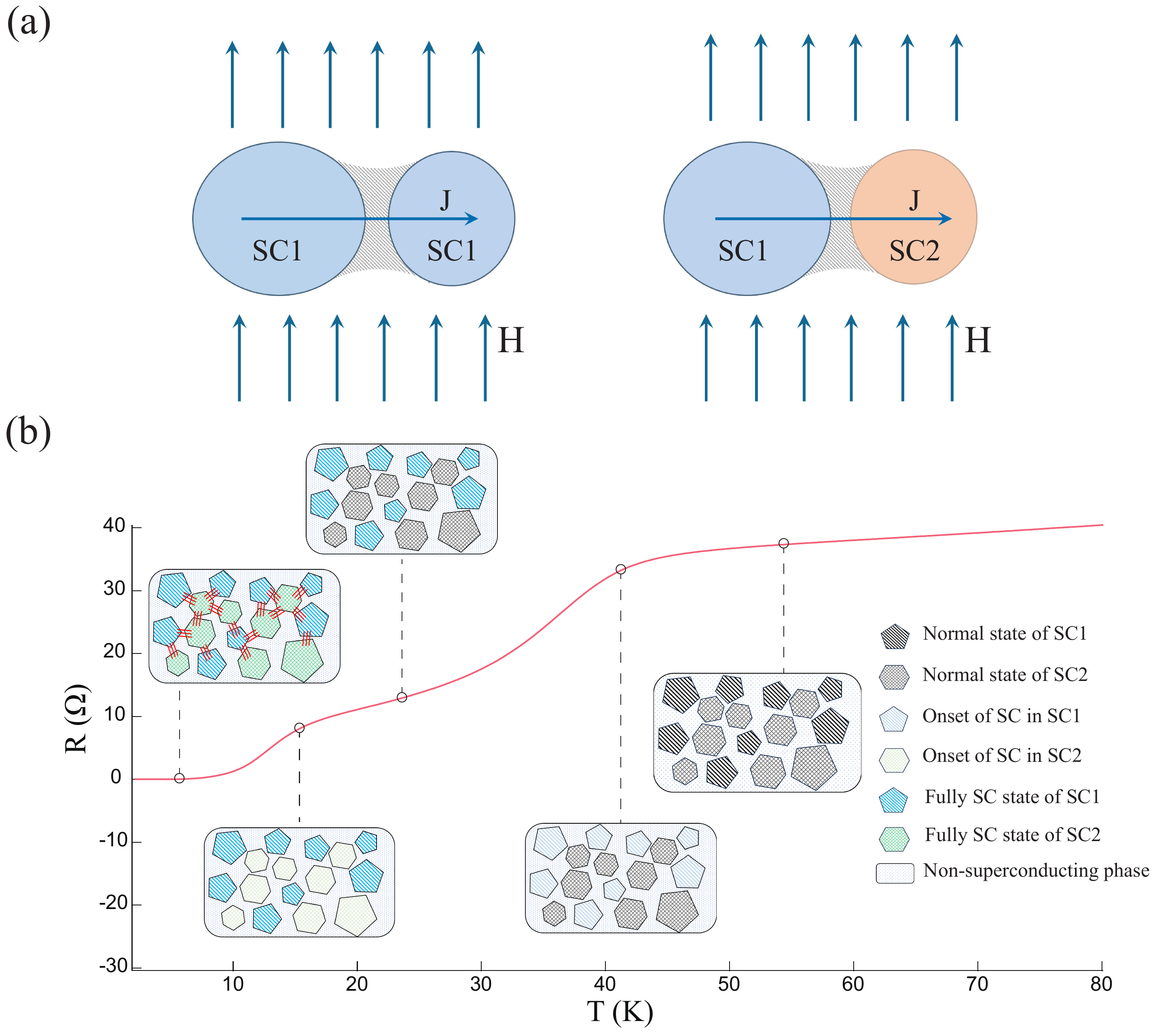}
\caption{\textbf{Schematic illustration of possible mechanisms underlying hysteresis in magnetoresistance and the second superconducting transition.} (a) Magnetic flux lines in superconducting grains (elliptical regions) and the intergranular medium (shaded regions) under an applied magnetic field. (b) Phase evolution of superconducting grains in two distinct phases, SC1 and SC2, with different $T_c$ values upon cooling. The superconducting phases and states are indicated by distinct shapes and colors as described in the legend. Red lines represent the formation of Josephson junctions, which allow dissipationless tunneling of current.}
\label{figure4}
\end{figure}

This phenomenon can be explained using the effective field model in granular superconductors\cite{balaev_correlation_2014}. As illustrated schematically in Fig.\ref{figure4}a, a granular superconductor consists of superconducting grains (elliptical regions) and non-superconducting weak link regions (shaded areas).
 In granular superconductors, the magnetoresistance hysteresis is often explained by an effective-field model. The weak‑link regions experience a field 
\begin{equation}
	B_{eff}(H)=H-4\pi M(H)\times\alpha
\end{equation}
where $\alpha$ accounts for demagnetizing factors of grains and the flux compression in the intergranular medium. On the ascending field branch, superconducting grains expel flux into the weak links, suppressing the critical current of the Josephson junctions and thereby increasing resistance. When the field surpasses the lower critical field $H_{c1}$, flux enters the grains. Upon reversing the field, pinned flux induces a paramagnetic moment, creating an induced field that opposes the external field. As a result, on the descending branch, the effective field is diminished, leading to a higher critical current and lower resistance at the same external field (Fig.\ref{figure3}b–g). The maximum cancellation between the effective field and the magnetization causes a resistance minimum at a positive field, and stably trapped flux within the grains explains the residual zero-field resistance R(0). The reduction of hysteresis with increasing temperature indicates the thermally activated nature of the Josephson network\cite{shifang_behaviour_1988}.

After establishing the granular nature of superconductivity in our films through transport measurements, we explain the origin of the two-step transitions within the Josephson-junction network model for granular superconductors\cite{quercia_physics_1984a}. As the temperature decreases, individual grains first become superconducting, developing a local order parameter amplitude while their phases can still fluctuate. The weak links between grains provide a Josephson coupling energy $E_J$. At higher temperatures, $k_B T \gg E_J$, the phases are uncorrelated, leading to finite resistance. When $E_J$ grows sufficiently with further cooling and becomes comparable to $k_B T$, phase correlations develop\cite{eley_approaching_2012}. This process is analogous to the Kosterlitz–Thouless transition in two-dimensional Josephson arrays\cite{resnick_kosterlitzthouless_1981a,abraham_resistive_1982a}, leading to the formation of phase-ordered clusters and, eventually, to long-range phase coherence below a second critical temperature $ T_{c,\text{onset}}^{\text{2nd}} $, at which the resistance vanishes.  

The effective field model described above for a granular superconductor adequately explains the observed two-step transitions and the details of the transport measurements. However, the nuanced findings—namely, a clear upturn in resistivity upon cooling call for a more refined framework. 
As shown in Fig.\ref{figure1}c, Film A exhibits two distinct resistivity upturns during the intermediate stage of ozone annealing, precisely aligning with the transition temperatures associated with the two-step transitions. These upturns cannot be simply attributed to the progressive formation of Josephson junction networks, as this mechanism would predict a monotonic decrease in resistance with decreasing temperature. Instead, the observed behavior strongly indicates more complex underlying processes at play. Resistance upturns are typically associated with superconducting fluctuations and result from disorder-induced charge localization or competing orders. Regardless of their causes, the appearance of a resistance upturn before both transitions indicates the presence of two distinct superconducting transition processes. Based on this experimental evidence, we propose an alternative scenario, illustrated in Fig.\ref{figure4}b. It indicates that two different superconducting grain phases can coexist within the same sample, each with its own transition temperature $T_c$: a high-$T_{c1}$ phase (SC1) and a low-$T_{c2}$ phase (SC2). The evolutionary process upon cooling can be described as follows. As the temperature decreases, the SC1 grains first undergo a superconducting transition, establishing internal phase coherence within each grain. When the temperature approaches $T_{c2}$ of SC2, the secondary transition occurs: SC2 grains become superconducting and similarly develop local phase coherence. Throughout this process, as an increasing number of grains enter the superconducting state, Josephson junctions progressively form at the weak links between grains (schematically represented by red lines in Fig.\ref{figure4}b). Once both SC1 and SC2 grains are superconducting, a connected Josephson-junction network is established, enabling the entire sample to reach a zero-resistance state.

It is important to note that the secondary transition region shows an extremely sensitive response to weak magnetic fields, as shown in Fig.\ref{figure2}. This trait arises from the inherent properties of Josephson junction networks. In granular superconductors, grain boundaries constitute Josephson-type weak links whose critical currents are highly responsive to applied magnetic fields\cite{balaev_hightemperature_2001,semenov_universal_2021,quercia_physics_1984a,eley_approaching_2012}. When these weak links form a network, the transport characteristics of the whole system can change significantly even in weak magnetic fields well below the intrinsic upper critical field of the grains\cite{balaev_hightemperature_2001,shifang_behaviour_1988}. Additionally, the thermally activated phase slippage mechanism provides additional insight into the resistive behavior of Josephson junctions at finite temperatures: an applied magnetic field adjusts the junction coupling energy, which greatly influences the thermal activation process, leading to transition broadening and field dependence of the secondary transition in the $\rho(T)$ curves\cite{ambegaokar_voltage_1969,tinkham_resistive_1988}. Therefore, even though we propose two types of superconducting grains with different critical temperatures, they remain coupled through the Josephson-junction network. This coupling naturally gives rise to the two key transport characteristics observed: the pronounced sensitivity to weak magnetic fields and the broadening of the secondary transition in the $\rho(T)$ curve.

The persistence of the secondary transition in thin films of bilayer nickelates likely stems from their tendency to lose oxygen. Even in samples that appear to be of high quality, where the resistivity drops to very low values near 30 K, a very weak secondary transition can still be discerned upon close inspection (see Fig.\ref{figure2} for Film B). The optimized preparation and ozone treatment significantly suppress this residual feature but do not eliminate it entirely. The clear direction forward offers promise for eventually suppressing the granular character and realizing homogeneous bulk superconductivity. The proposed scenario involving two types of superconducting grains bears some resemblances to the phase separation observed in superoxygenated La$_2$CuO$_{4+\delta}$, where excess oxygen leads to the coexistence of a superconducting and a magnetic stripe phase\cite{wells_incommensurate_1997}. A distinction in our case is that, in La$_2$PrNi$_2$O$_{7-\delta}$ thin films, both phases are superconducting and differ only in their transition temperatures. Finally, in contrast to the report in \cite{ji_timereversal_2026}, we find no evidence for a spin glass phase associated with the secondary transition in our samples. This discrepancy may be attributed to the different sample composition, notably the presence of Sm in their samples, which is absent from ours.

\section{conclusion}

In summary, we have achieved superconductivity in La$_2$PrNi$_2$O$_{7-\delta}$ thin films and identified their granular nature. The observed two-step superconducting transition, distinct magnetoresistance hysteresis, and sensitivity to weak magnetic fields are consistently explained by a model invoking the coexistence of two superconducting grain phases coupled through a Josephson-junction network. These findings provide new insight into the complex superconducting behavior of bilayer nickelate systems. Reducing oxygen inhomogeneity is essential for realizing bulk superconductivity with higher zero-resistance transition temperatures, which is crucial for reliable spectroscopic investigation of the superconducting mechanism.

\begin{acknowledgments}
This work was supported in part by the National Key Research and Development Program of China (Grants No. 2022YFA1403900 and No. 2021YFA1401800), the National Natural Science Foundation of China (Grant No. 12494593), Chinese Academy of Sciences (CAS) Superconducting Research Project (Grant No. SCZX-0101) and the Synergetic Extreme Condition User Facility (SECUF).
\end{acknowledgments}

\bibliography{reference.bib}

@article{liu_superconductivity_2025,
  title = {{Superconductivity and Normal-State Transport in Compressively Strained La$_{2}$PrNi$_{2}$O$_{7}$ Thin Films}},
  author = {Liu, Yidi and Ko, Eun Kyo and Tarn, Yaoju and Bhatt, Lopa and Li, Jiarui and Thampy, Vivek and Goodge, Berit H. and Muller, David A. and Raghu, Srinivas and Yu, Yijun and Hwang, Harold Y.},
  year = 2025,
  month = aug,
  journal = {Nat. Mater.},
  volume = {24},
  number = {8},
  pages = {1221--1227},
  issn = {1476-1122, 1476-4660},
  doi = {10.1038/s41563-025-02258-y},
  url = {https://www.nature.com/articles/s41563-025-02258-y},
  urldate = {2025-08-19},
  langid = {english},
  
}

@article{sun_signatures_2023,
	title = {{Signatures of superconductivity near 80 {K} in a nickelate under high pressure}},
	volume = {621},
	copyright = {2023 The Author(s), under exclusive licence to Springer Nature Limited},
	issn = {1476-4687},
	url = {https://www.nature.com/articles/s41586-023-06408-7},
	number = {7979},
	urldate = {2025-07-08},
	journal = {Nature},
	author = {Sun, Hualei and Huo, Mengwu and Hu, Xunwu and Li, Jingyuan and Liu, Zengjia and Han, Yifeng and Tang, Lingyun and Mao, Zhongquan and Yang, Pengtao and Wang, Bosen and Cheng, Jinguang and Yao, DaoXin and Zhang, GuangMing and Wang, Meng},
	month = sep,
	year = {2023},
	pages = {493--498},
	
}

@article{wang_bulk_2024,
  title = {{Bulk High-Temperature Superconductivity in Pressurized Tetragonal La$_{2}$PrNi$_{2}$O$_{7}$}},
  author = {Wang, Ningning and Wang, Gang and Shen, Xiaoling and Hou, Jun and Luo, Jun and Ma, Xiaoping and Yang, Huaixin and Shi, Lifen and Dou, Jie and Feng, Jie and Yang, Jie and Shi, Yunqing and Ren, Zhian and Ma, Hanming and Yang, Pengtao and Liu, Ziyi and Liu, Yue and Zhang, Hua and Dong, Xiaoli and Wang, Yuxin and Jiang, Kun and Hu, Jiangping and Nagasaki, Shoko and Kitagawa, Kentaro and Calder, Stuart and Yan, Jiaqiang and Sun, Jianping and Wang, Bosen and Zhou, Rui and Uwatoko, Yoshiya and Cheng, Jinguang},
  year = 2024,
  month = oct,
  journal = {Nature},
  volume = {634},
  number = {8034},
  pages = {579--584},
  issn = {0028-0836, 1476-4687},
  doi = {10.1038/s41586-024-07996-8},
  url = {https://www.nature.com/articles/s41586-024-07996-8},
  urldate = {2026-03-10},
  langid = {english},
  
}

@article{wang_pressureinduced_2024,
 title = {{Pressure-Induced Superconductivity In Polycrystalline La$_{2}$PrNi$_{2}$O$_{7-\delta}$}},
	volume = {14},
	url = {https://link.aps.org/doi/10.1103/PhysRevX.14.011040},
	doi = {10.1103/PhysRevX.14.011040},
	number = {1},
	journal = {Phys. Rev. X},
	author = {Wang, G. and Wang, N. N. and Shen, X. L. and Hou, J. and Ma, L. and Shi, L. F. and Ren, Z. A. and Gu, Y. D. and Ma, H. M. and Yang, P. T. and Liu, Z. Y. and Guo, H. Z. and Sun, J. P. and Zhang, G. M. and Calder, S. and Yan, J.Q. and Wang, B. S. and Uwatoko, Y. and Cheng, J.G.},
	month = mar,
	year = {2024},
	pages = {011040},
}

@article{wangwang_normal_2024,
  title = {{Normal and {{Superconducting Properties}} of La$_{3}$Ni$_{2}$O$_{7}$}},
  author = {Wang, Meng  and Wen , HaiHu  and Wu , Tao  and Yao , DaoXin  and Xiang , Tao },
  year = 2024,
  month = jul,
  journal = {Chin. Phys. Lett.},
  volume = {41},
  number = {7},
  pages = {077402},
  issn = {0256-307X, 1741-3540},
  doi = {10.1088/0256-307X/41/7/077402},
  url = {https://iopscience.iop.org/article/10.1088/0256-307X/41/7/077402},
  urldate = {2026-03-10},
  langid = {english},
  
}

@article{zhang_hightemperature_2024,
  title = {{High-Temperature Superconductivity with Zero Resistance and Strange-Metal Behaviour in {La$_{3}$Ni$_{2}$O$_{7-\delta}$}}},
  author = {Zhang, Yanan and Su, Dajun and Huang, Yanen and Shan, Zhaoyang and Sun, Hualei and Huo, Mengwu and Ye, Kaixin and Zhang, Jiawen and Yang, Zihan and Xu, Yongkang and Su, Yi and Li, Rui and Smidman, Michael and Wang, Meng and Jiao, Lin and Yuan, Huiqiu},
  year = 2024,
  month = aug,
  journal = {Nat. Phys.},
  volume = {20},
  number = {8},
  pages = {1269--1273},
  issn = {1745-2473, 1745-2481},
  doi = {10.1038/s41567-024-02515-y},
  url = {https://www.nature.com/articles/s41567-024-02515-y},
  urldate = {2026-03-10},
  langid = {english},
 
}

@article{zhou_superconductivity_2026,
  title = {{Superconductivity Onset above 60 {{K}} in Ambient-Pressure Nickelate Films}},
  author = {Zhou, Guangdi and Wang, Heng and Huang, Haoliang and Chen, Yaqi and Peng, Fei and Lv, Wei and Nie, Zihao and Wang, Wei and Jia, JinFeng and Xue, QiKun and Chen, Zhuoyu},
  year = 2026,
  month = mar,
  journal = {Natl. Sci. Rev.},
  pages = {nwag151},
  issn = {2095-5138, 2053-714X},
  doi = {10.1093/nsr/nwag151},
  url = {https://academic.oup.com/nsr/advance-article/doi/10.1093/nsr/nwag151/8512895},
  urldate = {2026-04-01},
  copyright = {https://creativecommons.org/licenses/by/4.0/},
  langid = {english},
  
}

@article{zhu_superconductivity_2024,
  title = {{Superconductivity in Pressurized Trilayer {La$_{4}$Ni$_{3}$O$_{10-\delta}$} Single Crystals}},
  author = {Zhu, Yinghao and Peng, Di and Zhang, Enkang and Pan, Bingying and Chen, Xu and Chen, Lixing and Ren, Huifen and Liu, Feiyang and Hao, Yiqing and Li, Nana and Xing, Zhenfang and Lan, Fujun and Han, Jiyuan and Wang, Junjie and Jia, Donghan and Wo, Hongliang and Gu, Yiqing and Gu, Yimeng and Ji, Li and Wang, Wenbin and Gou, Huiyang and Shen, Yao and Ying, Tianping and Chen, Xiaolong and Yang, Wenge and Cao, Huibo and Zheng, Changlin and Zeng, Qiaoshi and Guo, Jiangang and Zhao, Jun},
  year = 2024,
  month = jul,
  journal = {Nature},
  volume = {631},
  number = {8021},
  pages = {531--536},
  issn = {0028-0836, 1476-4687},
  doi = {10.1038/s41586-024-07553-3},
  url = {https://www.nature.com/articles/s41586-024-07553-3},
  urldate = {2026-03-10},
  langid = {english},
  
}

@article{houhou_emergence_2023,
  title = {{Emergence of High-Temperature Superconducting Phase in Pressurized La$_{3}$Ni$_{2}$O$_{7}$ Crystals}},
  author = {Hou, Jun and Yang, PengTao and Liu, ZiYi and Li, JingYuan and Shan, PengFei and Ma, Liang and Wang, Gang and Wang, NingNing and Guo, HaiZhong and Sun, JianPing and Uwatoko, Yoshiya and Wang, Meng and Zhang, GuangMing and Wang, BoSen and Cheng, JinGuang},
  year = {2023},
  month = {oct},
  journal = {Chin. Phys. Lett.},
  volume = {40},
  number = {11},
  pages = {117302},
  publisher = {{Chinese Physical Society and IOP Publishing Ltd}},
  doi = {10.1088/0256-307X/40/11/117302},
  url = {https://iopscience.iop.org/article/10.1088/0256-307X/40/11/117302},
}

@article{li_bulk_2026,
  title = {{Bulk Superconductivity up to 96 {{K}} in Pressurized Nickelate Single Crystals}},
  author = {Li, Feiyu and Xing, Zhenfang and Peng, Di and Dou, Jie and Guo, Ning and Ma, Liang and Zhang, Yulin and Wang, Lingzhen and Luo, Jun and Yang, Jie and Zhang, Jian and Chang, Tieyan and Chen, Yu-Sheng and Cai, Weizhao and Cheng, Jinguang and Wang, Yuzhu and Liu, Yuxin and Luo, Tao and Hirao, Naohisa and Matsuoka, Takahiro and Kadobayashi, Hirokazu and Zeng, Zhidan and Zheng, Qiang and Zhou, Rui and Zeng, Qiaoshi and Tao, Xutang and Zhang, Junjie},
  year = 2026,
  month = jan,
  journal = {Nature},
  volume = {649},
  number = {8098},
  pages = {871--878},
  issn = {0028-0836, 1476-4687},
  doi = {10.1038/s41586-025-09954-4},
  url = {https://www.nature.com/articles/s41586-025-09954-4},
  urldate = {2026-04-01},
  langid = {english},
  
}

@article{ko_signatures_2025a,
  title = {{Signatures of Ambient Pressure Superconductivity in Thin Film La$_{3}$Ni$_{2}$O$_{7}$}},
  author = {Ko, Eun Kyo and Yu, Yijun and Liu, Yidi and Bhatt, Lopa and Li, Jiarui and Thampy, Vivek and Kuo, ChengTai and Wang, Bai Yang and Lee, Yonghun and Lee, Kyuho and Lee, JunSik and Goodge, Berit H. and Muller, David A. and Hwang, Harold Y.},
  year = 2025,
  month = feb,
  journal = {Nature},
  volume = {638},
  number = {8052},
  pages = {935--940},
  issn = {0028-0836, 1476-4687},
  doi = {10.1038/s41586-024-08525-3},
  url = {https://www.nature.com/articles/s41586-024-08525-3},
  urldate = {2026-04-01},
  langid = {english},
  
}

@article{xiang_stabilizing_2026,
  title = {{Stabilizing and Tuning Superconductivity in La$_{3}$Ni$_{2}$O$_{7-\delta}$ Films: {{Oxygen}} Recycling Protocol Reveals Hole-Doping Analogue}},
  author = {Xiang, Lifen and Lei, Siyi and Ren, Xiaolin and Han, Ziao and Xu, Zijian and Zhou, X. J. and Zhu, Zhihai},
  year = 2026,
  month = mar,
  journal = {Phys. Rev. B},
  volume = {113},
  number = {10},
  pages = {104522},
  issn = {2469-9950, 2469-9969},
  doi = {10.1103/h9ls-y4s7},
  url = {https://link.aps.org/doi/10.1103/h9ls-y4s7},
  urldate = {2026-04-01},
  langid = {english},
  
}

@article{zhou_ambientpressure_2025a,
  title = {{Ambient-Pressure Superconductivity Onset above 40 {{K}} in (La,Pr)$_{3}$Ni$_{2}$O$_{7-\delta}$ Films}},
  author = {Zhou, Guangdi and Lv, Wei and Wang, Heng and Nie, Zihao and Chen, Yaqi and Li, Yueying and Huang, Haoliang and Chen, WeiQiang and Sun, YuJie and Xue, QiKun and Chen, Zhuoyu},
  year = 2025,
  month = apr,
  journal = {Nature},
  volume = {640},
  number = {8059},
  pages = {641--646},
  issn = {0028-0836, 1476-4687},
  doi = {10.1038/s41586-025-08755-z},
  url = {https://www.nature.com/articles/s41586-025-08755-z},
  urldate = {2026-04-01},
  langid = {english},
 
}

@article{li_angleresolved_2025,
  title = {{Angle-Resolved Photoemission Spectroscopy of Superconducting (La,Pr)$_{3}$Ni$_{2}$O$_{7}$/{{SrLaAlO$_{4}$}} Heterostructures}},
  author = {Li, Peng and Zhou, Guangdi and Lv, Wei and Li, Yueying and Yue, Changming and Huang, Haoliang and Xu, Lizhi and Shen, Jianchang and Miao, Yu and Song, Wenhua and Nie, Zihao and Chen, Yaqi and Wang, Heng and Chen, Weiqiang and Huang, Yaobo and Chen, Zhen-Hua and Qian, Tian and Lin, Junhao and He, Junfeng and Sun, Yu-Jie and Chen, Zhuoyu and Xue, QiKun},
  year = 2025,
  month = sep,
  journal = {Natl. Sci. Rev.},
  volume = {12},
  number = {10},
  pages = {nwaf205},
  issn = {2095-5138, 2053-714X},
  doi = {10.1093/nsr/nwaf205},
  url = {https://academic.oup.com/nsr/article/doi/10.1093/nsr/nwaf205/8140053},
  urldate = {2026-04-01},
  copyright = {https://creativecommons.org/licenses/by/4.0/},
  langid = {english},
  
}

@article{hao_superconductivity_2025,
  title = {{Superconductivity in {{Sr-doped La$_{3}$Ni$_{2}$O$_{7}$}} Thin Films}},
  author = {Hao, Bo and Wang, Maosen and Sun, Wenjie and Yang, Yang and Mao, Zhangwen and Yan, Shengjun and Sun, Haoying and Zhang, Hongyi and Han, Lu and Gu, Zhengbin and Zhou, Jian and Ji, Dianxiang and Nie, Yuefeng},
  year = 2025,
  month = nov,
  journal = {Nat. Mater.},
  volume = {24},
  number = {11},
  pages = {1756--1762},
  issn = {1476-1122, 1476-4660},
  doi = {10.1038/s41563-025-02327-2},
  url = {https://www.nature.com/articles/s41563-025-02327-2},
  urldate = {2026-04-01},
  langid = {english},
  
}

@article{lyu_preparation_2025,
  title = {{Preparation and Optimization of High-Temperature Superconducting {{Ruddlesden-Popper}} Nickelate Thin Films}},
  author = {Lyu, Wei and Nie, Zihao and Wang, Heng and Chen, Yaqi and Huang, Haoliang and Zhou, Guangdi and Xue, Qikun and Chen, Zhuoyu},
  year = 2025,
  journal = {Acta Phys. Sin.},
  volume = {74},
  number = {22},
  pages = {0},
  issn = {1000-3290, 1000-3290},
  doi = {10.7498/aps.74.20251080},
  url = {https://wulixb.iphy.ac.cn/article/doi/10.7498/aps.74.20251080},
  urldate = {2026-04-01},
  
}

@article{saito_highly_2016,
  title = {{Highly Crystalline {{2D}} Superconductors}},
  author = {Saito, Yu and Nojima, Tsutomu and Iwasa, Yoshihiro},
  year = 2016,
  month = dec,
  journal = {Nat. Rev. Mater.},
  volume = {2},
  number = {1},
  pages = {16094},
  issn = {2058-8437},
  doi = {10.1038/natrevmats.2016.94},
  url = {https://www.nature.com/articles/natrevmats201694},
  urldate = {2026-04-01},
  langid = {english},
  
}

@article{reyren_superconducting_2007,
  title = {{Superconducting {{Interfaces Between Insulating Oxides}}}},
  author = {Reyren, N. and Thiel, S. and Caviglia, A. D. and Kourkoutis, L. Fitting and Hammerl, G. and Richter, C. and Schneider, C. W. and Kopp, T. and R{\"u}etschi, A.S. and Jaccard, D. and Gabay, M. and Muller, D. A. and Triscone, J.M. and Mannhart, J.},
  year = 2007,
  month = aug,
  journal = {Science},
  volume = {317},
  number = {5842},
  pages = {1196--1199},
  issn = {0036-8075, 1095-9203},
  doi = {10.1126/science.1146006},
  url = {https://www.science.org/doi/10.1126/science.1146006},
  urldate = {2026-04-01},
  langid = {english},
  
}

@misc{ji_timereversal_2026,
  title = {{Time-Reversal Symmetry Breaking Superconductivity with Electronic Glass in Nickelate (La,Pr,Sm)$_{3}$Ni$_{2}$O$_{7}$ Films}},
  author = {Ji, Haoran and Xie, Zheyuan and Chen, Yaqi and Zhou, Guangdi and Pan, Longxin and Wang, Heng and Huang, Haoliang and Ge, Jun and Liu, Yi and Zhang, GuangMing and Wang, Ziqiang and Xue, QiKun and Chen, Zhuoyu and Wang, Jian},
  year = 2026,
  month = mar,
  number = {arXiv:2508.16412},
  eprint = {2508.16412},
  primaryclass = {cond-mat},
  publisher = {arXiv},
  doi = {10.48550/arXiv.2508.16412},
  url = {http://arxiv.org/abs/2508.16412},
  urldate = {2026-03-31},
  archiveprefix = {arXiv},
  
}

@article{saykin_spinglass_2025,
  title = {{Spin-Glass State in Nickelate Superconductors}},
  author = {Saykin, David R. and Gonzalez, Martin and Fowlie, Jennifer and Kivelson, Steven A. and Hwang, Harold Y. and Kapitulnik, Aharon},
  year = 2025,
  month = aug,
  journal = {npj Quantum Mater.},
  volume = {10},
  number = {1},
  pages = {94},
  issn = {2397-4648},
  doi = {10.1038/s41535-025-00813-z},
  url = {https://www.nature.com/articles/s41535-025-00813-z},
  urldate = {2026-04-01},
  langid = {english},
  
}

@article{luo_bilayer_2023,
  title = {{Bilayer {{Two-Orbital Model}} of La$_{3}$Ni$_{2}$O$_{7}$ under {{Pressure}}}},
  author = {Luo, Zhihui and Hu, Xunwu and Wang, Meng and W{\'u}, W{\'e}i and Yao, DaoXin},
  year = 2023,
  month = sep,
  journal = {Phys. Rev. Lett.},
  volume = {131},
  number = {12},
  pages = {126001},
  issn = {0031-9007, 1079-7114},
  doi = {10.1103/PhysRevLett.131.126001},
  url = {https://link.aps.org/doi/10.1103/PhysRevLett.131.126001},
  urldate = {2026-04-01},
  langid = {english}
}

@article{balaev_correlation_2014,
  title = {{Correlation {{Between Magnetoresistance}} and {{Magnetization Hysteresis}} in a {{Granular High-T C Superconductor}}: {{Impact}} of {{Flux Compression}} in the {{Intergrain Medium}}}},
  shorttitle = {Correlation {{Between Magnetoresistance}} and {{Magnetization Hysteresis}} in a {{Granular High-T C Superconductor}}},
  author = {Balaev, D. A. and Semenov, S. V. and Petrov, M. I.},
  year = 2014,
  month = jun,
  journal = {J Supercond. Nov. Magn.},
  volume = {27},
  number = {6},
  pages = {1425--1429},
  issn = {1557-1939, 1557-1947},
  doi = {10.1007/s10948-014-2491-6},
  url = {http://link.springer.com/10.1007/s10948-014-2491-6},
  urldate = {2026-01-05},
  copyright = {http://www.springer.com/tdm},
  langid = {english},
  
}

@article{balaev_magnetoresistance_2007,
  title = {{Magnetoresistance Hysteresis in Granular {{HTSCs}} as a Manifestation of the Magnetic Flux Trapped by Superconducting Grains in {{YBCO}} + {{CuO}} Composites}},
  author = {Balaev, D. A. and Gokhfeld, D. M. and Dubrovski{\u \i}, A. A. and Popkov, S. I. and Shaikhutdinov, K. A. and Petrov, M. I.},
  year = 2007,
  month = dec,
  journal = {J Exp. Theor. Phys.},
  volume = {105},
  number = {6},
  pages = {1174--1183},
  issn = {1063-7761, 1090-6509},
  doi = {10.1134/S1063776107120084},
  url = {http://link.springer.com/10.1134/S1063776107120084},
  urldate = {2026-01-28},
  copyright = {http://www.springer.com/tdm},
  langid = {english},
  
}

@article{shifang_behaviour_1988,
  title = {{The {{Behaviour}} of {{Negative Magnetoresistance}} and {{Hysteresis}} in {{YBa}}${_{2}}${{Cu}}${_{3}}${{O}}$_{{7-{\delta}}}$}},
  author = {Shifang, Sun and Yong, Zhao and Guoqiang, Pan and Daoqi, Yu and Han, Zhang and Zuyao, Chen and Yitai, Qian and Weiyan, Kuan and Qirui, Zhang},
  year = {1988},
  month = jun,
  journal = {Europhys. Lett.},
  volume = {6},
  number = {4},
  pages = {359--362},
  issn = {0295-5075, 1286-4854},
  doi = {10.1209/0295-5075/6/4/014},
  url = {https://iopscience.iop.org/article/10.1209/0295-5075/6/4/014},
  urldate = {2026-01-26},
  langid = {english},
  
}

@article{ji_magneticfielddependent_1993,
  title = {{Magnetic-Field-Dependent Surface Resistance and Two-Level Critical-State Model for Granular Superconductors}},
  author = {Ji, L. and Rzchowski, M. S. and Anand, N. and Tinkham, M.},
  year = 1993,
  month = jan,
  journal = {Phys. Rev. B},
  volume = {47},
  number = {1},
  pages = {470--483},
  issn = {0163-1829, 1095-3795},
  doi = {10.1103/PhysRevB.47.470},
  url = {https://link.aps.org/doi/10.1103/PhysRevB.47.470},
  urldate = {2026-01-05},
  copyright = {http://link.aps.org/licenses/aps-default-license},
  langid = {english},
  
}

@article{quercia_physics_1984a,
  title = {{Physics and Applications of the {{Josephson}} Effect}},
  author = {A. Barone and G. Paterno},
  year = 1984,
  month = oct,
  journal = {Il Nuovo Cimento D},
  volume = {4},
  number = {4},
  pages = {411--412},
  issn = {1826-9893},
  doi = {10.1007/BF02451297},
  url = {https://doi.org/10.1007/BF02451297},
  urldate = {2026-04-02},
  copyright = {http://www.springer.com/tdm},
  langid = {english}
}

@article{resnick_kosterlitzthouless_1981a,
  title = {{Kosterlitz-{{Thouless Transition}} in {{Proximity-Coupled Superconducting Arrays}}}},
  author = {Resnick, D. J. and Garland, J. C. and Boyd, J. T. and Shoemaker, S. and Newrock, R. S.},
  year = 1981,
  month = nov,
  journal = {Phys. Rev. Lett.},
  volume = {47},
  number = {21},
  pages = {1542--1545},
  issn = {0031-9007},
  doi = {10.1103/PhysRevLett.47.1542},
  url = {https://link.aps.org/doi/10.1103/PhysRevLett.47.1542},
  urldate = {2026-04-02},
  copyright = {http://link.aps.org/licenses/aps-default-license},
  langid = {english}
}

@article{abraham_resistive_1982a,
  title = {{Resistive Transition in Two-Dimensional Arrays of Superconducting Weak Links}},
  author = {Abraham, David W. and Lobb, C. J. and Tinkham, M. and Klapwijk, T. M.},
  year = 1982,
  month = nov,
  journal = {Phys. Rev. B},
  volume = {26},
  number = {9},
  pages = {5268--5271},
  issn = {0163-1829},
  doi = {10.1103/PhysRevB.26.5268},
  url = {https://link.aps.org/doi/10.1103/PhysRevB.26.5268},
  urldate = {2026-04-02},
  copyright = {http://link.aps.org/licenses/aps-default-license},
  langid = {english},
 
}

@article{balaev_hightemperature_2001,
  title = {{High-Temperature Superconductor Based Composites: {{Large}} Magnetoresistance in Weak Magnetic Fields}},
  shorttitle = {High-Temperature Superconductor Based Composites},
  author = {Balaev, D. A. and Gohfeld, D. M. and Popkov, S. I. and Saihutdinov, K. A. and Petrov, M. I.},
  year = 2001,
  month = nov,
  journal = {Tech. Phys. Lett.},
  volume = {27},
  number = {11},
  pages = {952--955},
  issn = {1063-7850, 1090-6533},
  doi = {10.1134/1.1424404},
  url = {http://link.springer.com/10.1134/1.1424404},
  urldate = {2026-02-03},
  copyright = {http://www.springer.com/tdm},
  langid = {english},
  
}

@article{semenov_universal_2021,
  title = {{Universal {{Behavior}} and {{Temperature Evolution}} of the {{Magnetoresistance Hysteresis}} in {{Granular High-Temperature Superconductors Y}}--{{Ba}}--{{Cu}}--{{O}}}},
  author = {Semenov, S. V. and Balaev, D. A. and Petrov, M. I.},
  year = 2021,
  month = jul,
  journal = {Phys. Solid State},
  volume = {63},
  number = {7},
  pages = {1069--1080},
  issn = {1063-7834, 1090-6460},
  doi = {10.1134/S1063783421070192},
  url = {https://link.springer.com/10.1134/S1063783421070192},
  urldate = {2026-02-04},
  langid = {english},
  
}

@article{eley_approaching_2012,
  title = {{Approaching Zero-Temperature Metallic States in Mesoscopic Superconductor--Normal--Superconductor Arrays}},
  author = {Eley, Serena and Gopalakrishnan, Sarang and Goldbart, Paul M. and Mason, Nadya},
  year = 2012,
  month = jan,
  journal = {Nat. Phys.},
  volume = {8},
  number = {1},
  pages = {59--62},
  issn = {1745-2473, 1745-2481},
  doi = {10.1038/nphys2154},
  url = {https://www.nature.com/articles/nphys2154},
  urldate = {2026-02-23},
  langid = {english},
  
}

@article{ambegaokar_voltage_1969,
  title = {{Voltage {{Due}} to {{Thermal Noise}} in the Dc {{Josephson Effect}}}},
  author = {Ambegaokar, Vinay and Halperin, B. I.},
  year = 1969,
  month = jun,
  journal = {Phys. Rev. Lett.},
  volume = {22},
  number = {25},
  pages = {1364--1366},
  issn = {0031-9007},
  doi = {10.1103/PhysRevLett.22.1364},
  url = {https://link.aps.org/doi/10.1103/PhysRevLett.22.1364},
  urldate = {2026-02-04},
  copyright = {http://link.aps.org/licenses/aps-default-license},
  langid = {english},
  
}

@article{tinkham_resistive_1988,
  title = {{Resistive {{Transition}} of {{High-Temperature Superconductors}}}},
  author = {Tinkham, M.},
  year = 1988,
  month = oct,
  journal = {Phys. Rev. Lett.},
  volume = {61},
  number = {14},
  pages = {1658--1661},
  issn = {0031-9007},
  doi = {10.1103/PhysRevLett.61.1658},
  url = {https://link.aps.org/doi/10.1103/PhysRevLett.61.1658},
  urldate = {2026-04-02},
  copyright = {http://link.aps.org/licenses/aps-default-license},
  langid = {english},
  
}

@misc{wang_electronic_2025,
  title = {{Electronic Structure of Compressively Strained Thin Film La$_{2}$PrNi$_{2}$O$_{7}$}},
  author = {Wang, Bai Yang and Zhong, Yong and Abadi, Sebastien and Liu, Yidi and Yu, Yijun and Zhang, Xiaoliang and Wu, Yi-Ming and Wang, Ruohan and Li, Jiarui and Tarn, Yaoju and Ko, Eun Kyo and Thampy, Vivek and Hashimoto, Makoto and Lu, Donghui and Lee, Young S. and Devereaux, Thomas P. and Jia, Chunjing and Hwang, Harold Y. and Shen, Zhi-Xun},
  year = 2025,
  publisher = {arXiv},
  number = {arXiv:2504.16372},
  eprint = {2504.16372},
  primaryclass = {cond-mat},
  doi = {10.48550/arXiv.2504.16372},
  url = {https://arxiv.org/abs/2504.16372},
  urldate = {2026-04-02},
  copyright = {arXiv.org perpetual, non-exclusive license}
}

@misc{sun_observation_2025,
  title = {{Observation of Superconductivity-Induced Leading-Edge Gap in {{Sr-doped}} La$_{3}$Ni$_{2}$O$_{7}$ Thin Films}},
  author = {Sun, Wenjie and Jiang, Zhicheng and Hao, Bo and Yan, Shengjun and Zhang, Hongyi and Wang, Maosen and Yang, Yang and Sun, Haoying and Liu, Zhengtai and Ji, Dianxiang and Gu, Zhengbin and Zhou, Jian and Shen, Dawei and Feng, Donglai and Nie, Yuefeng},
  year = 2025,
  month = jul,
  number = {arXiv:2507.07409},
  eprint = {2507.07409},
  primaryclass = {cond-mat},
  publisher = {arXiv},
  doi = {10.48550/arXiv.2507.07409},
  url = {http://arxiv.org/abs/2507.07409},
  urldate = {2026-04-02},
  archiveprefix = {arXiv},
  
}

@misc{shen_nodeless_2025,
  title = {{Nodeless Superconducting Gap and Electron-Boson Coupling in (La,Pr,Sm)$_{3}$Ni$_{2}$O$_{7}$ Films}},
  author = {Shen, Jianchang and Zhou, Guangdi and Miao, Yu and Li, Peng and Ou, Zhipeng and Chen, Yaqi and Wang, Zechao and Luan, Runqing and Sun, Hongxu and Feng, Zikun and Yong, Xinru and Li, Yueying and Xu, Lizhi and Lv, Wei and Nie, Zihao and Wang, Heng and Huang, Haoliang and Sun, Yu-Jie and Xue, Qi-Kun and He, Junfeng and Chen, Zhuoyu},
  year = 2025,
  publisher = {arXiv},
  doi = {10.48550/ARXIV.2502.17831},
  eprint = {2502.17831},
  primaryclass = {cond-mat},
  url = {https://arxiv.org/abs/2502.17831},
  urldate = {2026-04-03},
  copyright = {arXiv.org perpetual, non-exclusive license}
}

@article{zhang_bulk_2025,
  title = {{Bulk Superconductivity in Pressurized Trilayer Nickelate Pr$_{4}$Ni$_{3}$O$_{10}$ Single Crystals}},
  author = {Zhang, Enkang and Peng, Di and Zhu, Yinghao and Chen, Lixing and Cui, Bingkun and Wang, Xingya and Wang, Wenbin and Zeng, Qiaoshi and Zhao, Jun},
  year = 2025,
  month = apr,
  journal = {Phys. Rev. X},
  volume = {15},
  number = {2},
  pages = {021008},
  issn = {2160-3308},
  doi = {10.1103/PhysRevX.15.021008},
  url = {https://link.aps.org/doi/10.1103/PhysRevX.15.021008},
  urldate = {2026-04-03},
  langid = {english}
}

@article{shi_pressure_2025,
  title = {{Pressure Induced Superconductivity in Hybrid {{Ruddlesden}}-{{Popper La${_5}$Ni${_3}$O${_{11}}$}} Single Crystals}},
  author = {Shi, Mengzhu and Peng, Di and Fan, Kaibao and Xing, Zhenfang and Yang, Shaohua and Wang, Yuzhu and Li, Houpu and Wu, Rongqi and Du, Mei and Ge, Binghui and Zeng, Zhidan and Zeng, Qiaoshi and Ying, Jianjun and Wu, Tao and Chen, Xianhui},
  year = 2025,
  month = nov,
  journal = {Nat. Phys.},
  volume = {21},
  number = {11},
  pages = {1780--1786},
  issn = {1745-2473, 1745-2481},
  doi = {10.1038/s41567-025-03023-3},
  url = {https://www.nature.com/articles/s41567-025-03023-3},
  urldate = {2026-04-03},
  langid = {english}
}

@article{shi_critical_2026,
  title = {{Critical {{Thickness}} and {{Long}}-{{Term Ambient Stability}} in Superconducting LaPr$_{2}$Ni$_{2}$O$_{7}$ {{Films}}}},
  author = {Shi, Yuexin and Song, Chenyao and Jia, Yingze and Wang, Yanzhi and Li, Qi and Chen, Ye and Yang, Yue and Fu, Junchi and Qin, Ming and Song, Dongsheng and Chen, Zhen and Yuan, Huiqiu and Xie, Yanwu and Zhang, Meng},
  year = 2026,
  month = jan,
  journal = {Adv. Mater.},
  volume = {38},
  number = {4},
  pages = {e10394},
  issn = {0935-9648, 1521-4095},
  doi = {10.1002/adma.202510394},
  url = {https://advanced.onlinelibrary.wiley.com/doi/10.1002/adma.202510394},
  urldate = {2026-04-03},
  langid = {english}
}

@article{wells_incommensurate_1997,
  title = {{Incommensurate {{Spin Fluctuations}} in {{High-Transition Temperature Superconductors}}}},
  author = {Wells, B. O. and Lee, Y. S. and Kastner, M. A. and Christianson, R. J. and Birgeneau, R. J. and Yamada, K. and Endoh, Y. and Shirane, G.},
  year = 1997,
  month = aug,
  journal = {Science},
  volume = {277},
  number = {5329},
  pages = {1067--1071},
  issn = {0036-8075, 1095-9203},
  doi = {10.1126/science.277.5329.1067},
  url = {https://www.science.org/doi/10.1126/science.277.5329.1067},
  urldate = {2026-04-03},
  langid = {english}
}

@article{ren_resolving_2025a,
  title = {{Resolving the Electronic Ground State of {{La$_3$Ni$_2$O$_{7-\delta}$}} Films}},
  author = {Ren, Xiaolin and Sutarto, Ronny and Wu, Xianxin and Zhang, Jianfeng and Huang, Hai and Xiang, Tao and Hu, Jiangping and Comin, Riccardo and Zhou, Xingjiang and Zhu, Zhihai},
  year = 2025,
  month = feb,
  journal = {Commun. Phys.},
  volume = {8},
  number = {1},
  pages = {52},
  issn = {2399-3650},
  doi = {10.1038/s42005-025-01971-z},
  url = {https://www.nature.com/articles/s42005-025-01971-z},
  urldate = {2025-11-08},
  langid = {english}
}

\end{document}